\documentclass[twocolumn,showpacs,preprintnumbers,amsmath,amssymb]{revtex4}

\usepackage{graphicx}
\usepackage{dcolumn}
\usepackage{bm}

\begin{document}

\title{Information Filtering via Self-Consistent Refinement}
\author{Jie Ren$^{1,2}$}
\author{Tao Zhou$^{1,3,4}$}
\email{zhutou@ustc.edu}
\author{Yi-Cheng Zhang$^{1,4}$}

\affiliation{%
$^1$ Department of Physics, University of Fribourg, Chemin du Muse
3, 1700 Fribourg Switzerland \\ $^2$ Department of Physics and
Centre for Computational Science and Engineering, National
University of Singapore, Singapore 117542, Republic of Singapore
\\ $^3$ Department of Modern Physics, University of Science and Technology of China, Hefei 230026, PR
China \\ $^4$ Information Economy and Internet Research Laboratory,
University of Electronic Science and Technology of
China, Chengdu 610054, PR China }%

\begin{abstract}
Recommender systems are significant to help people deal with the
world of information explosion and overload. In this Letter, we
develop a general framework named self-consistent refinement and
implement it be embedding two representative recommendation
algorithms: similarity-based and spectrum-based methods. Numerical
simulations on a benchmark data set demonstrate that the present
method converges fast and can provide quite better performance than
the standard methods.
\end{abstract}


\maketitle

\emph{Introduction}.---The last few years have witnessed an
explosion of information that the Internet and World Wide Web have
brought us into a world of endless possibilities: people may choose
from thousands of movies, millions of books, and billions of web
pages. The amount of information is increasing more quickly than our
processing ability, thus evaluating all these alternatives and then
making choice becomes infeasible. As a consequence, an urgent
problem is how to automatically extract the hidden information and
do a personal recommendation. For example, Amazon.com uses one's
purchase record to recommend books \cite{Linden2003}, and
AdaptiveInfo.com uses one's reading history to recommend news
\cite{Billsus2002}. Motivated by the significance in economy and
society, the design of an efficient recommendation algorithm becomes
a joint focus from engineering science
\cite{Herlocker2004,Adomavicius2005} to marketing practice
\cite{Ansari2000,Ying2006}, from mathematical analysis
\cite{Kumar2001,Donovan2005} to physics community
\cite{Maslov2001,Laureti2006,Zhang2007a,Zhang2007b,Zhou2007a,Zhou2007b,Yu2006,Blattner2007}.

A recommender system, consisted of $N$ users and $M$ items, can be
fully described by an $N\times M$ rating matrix $R$, with
$R_{i\alpha}\neq 0$ the rating user $i$ gives to item $\alpha$. If
$i$ has not yet evaluated $\alpha$, $R_{i\alpha}$ is set as zero.
The aim of a recommender system, or of a recommendation algorithm,
is to predict ratings for the items have not been voted. To evaluate
the algorithmic accuracy, the given data set is usually divided into
two parts: one is the training set, and the other one is the testing
set. Only the information contained in the training set can be used
in the prediction. Denoting the predicted rating matrix as
$\tilde{R}$, the most commonly used measurement for the algorithmic
accuracy, namely the \emph{mean average error} (MAE), is defined as:
\begin{equation}
\emph{MAE}=\frac{1}{S}\sum_{(i,\alpha)}|\tilde{R}_{i\alpha}-R^*_{i\alpha}|,
\end{equation}
where the subscript $(i,\alpha)$ runs over all the elements
corresponding to the non-zero ratings in testing set, $R^*$ denotes
the rating matrix for testing set, and $S$ is the number of non-zero
ratings in $R^*$.

Thus far, the most accurate algorithms are content-based
\cite{Pazzani2007}. However, those methods are practical only if the
items have well-defined attributes, and those attributes can be
extracted automatically. Besides the content-based algorithms, the
recommendation methods can be classified into two main categories:
similarity-based \cite{Konstan1997,Sarwar2001} and spectrum-based
\cite{Billsus1998,Sarwar2000}. In this Letter, we propose a generic
framework of \emph{self-consistent refinement} (SCR) for the
personal recommendation, which is implemented by embedding the
similarity-based and spectrum-based methods, respectively. Numerical
simulations on a benchmark data set demonstrate the significant
improvement of algorithmic performance via SCR compared with the
standard methods.

\emph{Generic framework of SCR}.---The similarity-based and
spectrum-based algorithms, including their extensions, can be
expressed in a generic matrix formula
\begin{equation}
\tilde{R}=\mathfrak{D}(R),
\end{equation}
where $R$ is the rating matrix obtained from the training set,
$\tilde{R}$ the predicted rating matrix, and $\mathfrak{D}$ a matrix
operator. This operator, $\mathfrak{D}$, may be extremely simple as
a left-multiplying matrix used in the basic similarity-based method,
or very complicated, usually involving a latent optimization
process, like the case of rank-$k$ singular value decomposition (see
below for details). Most previous works concentrated on the design
of the operator $\mathfrak{D}$. In contrast, we propose a completely
new scenario where Eq. (2) is replaced by a SCR via iterations.
Denoting the initial configuration $R^{(0)}=R$, and the initial time
step $k=0$, a generic framework of SCR reads:

(i) Implement the operation $\mathfrak{D}(R^{(k)})$;

(ii) Set the elements of $R^{(k+1)}$ as
\begin{equation}
    R_{i\alpha}^{(k+1)}=\left\{
    \begin{array}{cc}
\mathfrak{D}(R^{(k)})_{i\alpha}, & R_{i\alpha}=0, \\
R_{i\alpha}, & R_{i\alpha}\neq 0.
    \end{array}
    \right.
\end{equation}
Then, set $k=k+1$.

(iii) Repeat (i)(ii) until the difference between $R^{(k)}$ and
$R^{(k-1)}$ (or, more practical, the difference $| MAE(k) - MAE(k-1)
|$) is smaller than a given terminate threshold.

Consider the matrix series $R^{(0)},R^{(1)},\cdots,R^{(T)}$ ($T$
denotes the last time step) as a certain dynamics driven by the
operator $\mathfrak{D}$, all the elements corresponding to the voted
items (i.e. $R_{i\alpha}\neq 0$) can be treated as the
\emph{boundary conditions} giving expression to the known
information. If $\tilde{R}$ is an ideal prediction, consider itself
as the known rating matrix, is should satisfy the self-consistent
condition $\tilde{R}= \mathfrak{D}(\tilde{R})$. However, this
equation is not hold for the standard methods. Correspondingly, the
convergent matrix $R^{(T)}$ is self-consistent. Though the
simplicity of SCR, it leads to a great improvement compared with the
traditional case shown in Eq. (2).

\emph{Similarity-based SCR}.---The basic idea behind the
similarity-based method is that: a user who likes a item will also
like other similar items \cite{Sarwar2001}. Taking into account the
different evaluation scales of different users
\cite{Zhang2007b,Blattner2007}, we subtract the corresponding user
average from each evaluated entry in the matrix $R$ and get a new
matrix $R'$. The similarity between items $\alpha$ and $\beta$ is
given by:
\begin{equation}
\Omega_{\alpha\beta}=\frac{\sum_{i\in U}R'_{i\alpha}\cdot
R'_{i\beta}}{\sqrt{\sum_{i\in U}R_{i\alpha}^{'2}}\sqrt{\sum_{i\in
U}R_{i\beta}^{'2}}}\in[-1, 1],
\end{equation}
where $\langle R\rangle_i$ is the average evaluation of user $i$ and
$R'_{i\alpha}=R_{i\alpha}-\langle R\rangle_i$. $U$ denotes the set
of users who evaluated both items $\alpha$ and $\beta$.
$\Omega_{\alpha\beta}\rightarrow 1$ means the items $\alpha$ and
$\beta$ are very similar, while $\Omega_{\alpha\beta}\rightarrow -1$
means the opposite case.

In the most widely applied similarity-based algorithm, namely
\emph{collaborative filtering} \cite{Resnick1994,Breese1998}, the
predicted rating is calculated by using a weighted average, as:
\begin{equation}
\tilde{R}_{i\alpha}=\frac{\sum_{\beta}\Omega_{\alpha\beta}\cdot
R'_{i\beta}}{\sum_{\beta}|\Omega_{\alpha\beta}|}.
\end{equation}
The contribution of $\Omega_{\alpha\beta}\cdot R'_{i\beta}$ is
positive if the signs of $\Omega_{\alpha\beta}$ and $R'_{i\beta}$
are the same. That is to say, a person $i$ like item $\alpha$ may
result from the situations (i) the person $i$ likes the item $\beta$
which is similar to item $\alpha$, or (ii) the person $i$ dislikes
the item $\beta$ which is opposite to item $\alpha$ (i.e.
$\Omega_{\alpha\beta}<0$). Note that, when computing the predictions
to a specific user $i$, we have to add the average rating of this
user, $\langle R\rangle_i$, back to $\tilde{R}_{i\alpha}$.

Obviously, Eq. (5) can be rewritten in a matrix form for any given
user $i$, as
\begin{equation}
\tilde{R_i}=P\cdot R'_i,
\end{equation}
where $\tilde{R_i}$ and $R'_i$ are $M$-dimensional column vectors
denoting the predicted and known ratings for user $i$, and
$P=\sum_{\beta}\Omega_{\alpha\beta}/\sum_{\beta}|\Omega_{\alpha\beta}|$,
acting as the transfer matrix. For simplicity, hereinafter, without
confusion, we cancel the subscript $i$ and superscript - a comma.
Since for each user, the predicting operation can be expressed in a
matrix form, we can get the numerical results by directly using the
general framework of SCR, as shown in Eq. (3). However, we have to
perform the matrix multiplying for every user, which takes long time
in computation especially for huge-size recommender systems.

To get the analytical expression and reduce the computational
complexity, for a given user, we group its known ratings (as
boundary conditions) and unknown ratings into $R_B$ and $R_U$,
respectively. Correspondingly, matrix $P$ is re-arranged by the same
order as $R$. For this user, we can rewrite Eq. (6) in a sub-matrix
multiplying form:
\begin{equation}
\label{Laplacian}
\begin{pmatrix}
\tilde{R}_{B} \\
\tilde{R}_{U}
\end{pmatrix} =
\begin{pmatrix}
P_{BB} & P_{BU}\\
P_{UB} & P_{UU}
\end{pmatrix}
\begin{pmatrix}
R_{B} \\
R_{U}
\end{pmatrix}.
\end{equation}
In the standard collaborative filtering
\cite{Resnick1994,Breese1998}, as shown in Eq. (5), the unknown
vector, $R_U$, is set as a zero vector. Therefore, the predicted
vector, $\tilde{R}_U$, can be expressed by a compact form:
\begin{equation}
\tilde{R}_U=P_{UB}\cdot R_B.
\end{equation}
Clearly, it only takes into account the direct correlations between
the unknown and known sets.

The solution Eq. (8) does not obey the self-consistent condition,
for the free sub-vector $\tilde{R}_U$ is not equal to $R_U$.
Considering the self-consistent condition (i.e $\tilde{R}_U=R_U$),
the predicted vector should obey the following equation:
\begin{equation}
\tilde{R}_U=P_{UB}R_B+P_{UU}\tilde{R}_U,
\end{equation}
whose solution reads:
\begin{equation}
\tilde{R}_U=(I-P_{UU})^{-1}P_{UB}R_B.
\end{equation}
This solution differs from the standard collaborative filtering by
an additional item $(I-P_{UU})^{-1}$.

Since it may not be practical to directly inverse $(I-P_{UU})$
especially for huge-size $P_{UU}$, we come up with a simple and
efficient iterative method: Substitute the first results
$\tilde{R}_U$ for $R_U$, on the right term of Eq. (6), and take
$R_B$ as the fixed boundary conditions. Then, get the second step
results about $\tilde{R}_U$, and substitute it for $R_U$ again. Do
it repeatedly, at the $n$th step, we get:
\begin{equation}
\tilde{R}_U=(I+P_{UU}+P^2_{UU}+\cdots+P^{n-1}_{UU})P_{UB}R_B.
\end{equation}
Since the dominant eigenvalue of $P_{UU}$ is smaller than $1$,
$P_{UU}^n$ converges exponentially fast \cite{Golub1996}, and we can
get the stable solution quickly within several steps.

In addition, besides the item-item similarity used introduced here,
the similarity-based method can also be implemented analogously via
using the user-user similarity \cite{Adomavicius2005}. The SCR can
also be embedded in that case, and gain much better algorithmic
accuracy.

\begin{figure}
\scalebox{0.6}[0.6]{\includegraphics{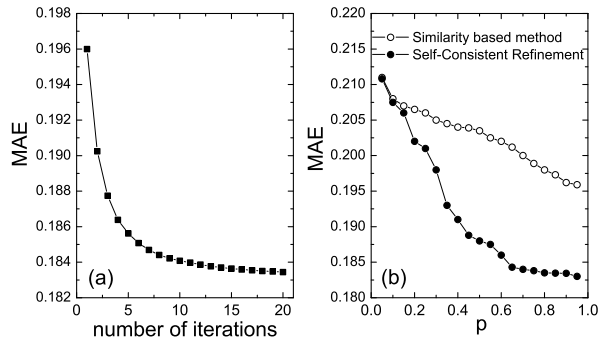}} \caption{(a)
Prediction error vs. iteration step, with $p=0.9$ fixed. (b) The
comparison of algorithmic accuracy between the standard
similarity-based method and the similarity-based SCR for different
$p$.}
\end{figure}

\emph{Spectrum-based SCR}.---We here present a spectrum-based
algorithm, which relies on the Singular Value Decomposition (SVD) of
the rating matrix. Analogously, we use the matrix with subtraction
of average ratings, $R'$, instead of $R$. The SVD of $R'$ is defined
as \cite{Zhang2004}:
\begin{equation}
R'=U\cdot S\cdot V^T,
\end{equation}
where $U$ is an $N\times N$ unitary matrix formed by the
eigenvectors of $R'R'^T$, $S$ is an $N\times M$ singular value
matrix with nonnegative numbers in decreasing order on the diagonal
and zeros off the diagonal, and $V^T$ is an $M\times M$ unitary
matrix formed by the eigenvectors of $R'^TR'$. The number of
positive diagonal elements in $S$ equals $\texttt{rank}(R')$.

We keep only the $k$ largest diagonal elements (also the $k$ largest
singular values) to obtain a reduced $k\times k$ matrix $S_k$, and
then, reduce the matrices $U$ and $V$ accordingly. That is to say,
only the $k$ column vectors of $U$ and $k$ row vectors of $V^T$
corresponding to the $k$ largest singular values are kept. The
reconstructed matrix reads:
\begin{equation}
R'_k=U_k\cdot S_k \cdot V^T_k,
\end{equation}
where $U_k$, $S_k$ and $V^T_k$ have dimensions $N\times k$, $k\times
k$ and $k\times M$, respectively. Note that, Eq. (13) is no longer
the exact decomposition of the original matrix $R'$ (i.e., $R'_k\neq
R'$), but the closest rank-$k$ matrix to $R$ \cite{Horn1985}. In
other words, $R'_k$ minimizes the Frobenius norm $\|R'-R'_k\|$
\cite{norm} over all rank-$k$ matrices. Previous studies found that
\cite{Berry1995} the reduced dimensional approximation sometimes
performs better than the original matrix in information retrieval
since it filters out the small singular values that may be highly
distorted by the noise.

Actually, each row of the $N\times k$ matrix $U_k\sqrt{S_k}$
represents the vector of the corresponding agent's tastes, and each
row of the $M\times k$ matrix $V_k\sqrt{S_k}$ characterizes the
features of the corresponding item. Therefore, the prediction of the
evaluation a user $i$ gives to an item $\alpha$ can be obtained by
computing the inner product of the $i$-th row of $U_k\sqrt{S_k}$ and
the $\alpha$-th row of $V_k\sqrt{S_k}$:
\begin{eqnarray}
\tilde{R}&=&U_k\sqrt{S_k}\cdot(V_k\sqrt{S_k})^T=U_k\sqrt{S_k}\cdot\sqrt{S_k}V_k^T\nonumber \\
&=&U_k\cdot S_k \cdot V^T_k=R_k.
\end{eqnarray}
This derivation reproduces the Eq. (13), and illuminates the reason
why using SVD to extract hidden information in user-item rating
matrix. The entry $\tilde{R}_{i\alpha}$ is the predicted rating of
user $i$ on item $\alpha$.

An underlying assumption in the $k$-truncated SVD method is the
existence of $k$ principle attributes in both the user's tastes and
the item's features. For example, a movie's attributes may include
the director, hero, heroine, gut, music, etc., and a user has his
personal tastes on each attribute. If a movie is well fit his
tastes, he will give a high rating, otherwise a low rating. Denote
the states of a user $i$ and an item $\alpha$ as:
\begin{equation}
\langle u_i|=(u^1_i,u^2_i,\cdots,u^k_i); \langle
v_{\alpha}|=(v^1_{\alpha},v^2_{\alpha},\cdots,v^k_{\alpha}),
\end{equation}
then we can estimate the evaluation of $i$ on $\alpha$ as the
matching extent between their tastes and features:
\begin{equation}
\tilde{R}_{i\alpha}=\langle u_i|v_{\alpha}\rangle.
\end{equation}
Therefore, we want to find a matrix $\tilde{R}$ that can be
decomposed to $N$ $k$-dimensional taste vectors and $M$
$k$-dimensional feature vectors so that the corresponding entries
are exactly the same as the known ratings and consequently, the
other entries are the predicted ratings.

However, the $k$-truncated SVD matrix is not self-consistent for the
elements corresponding to the known ratings in $R'_k$ are not
exactly the same as those in $R'$. A self-consistent prediction
matrix can be obtained via an iterative $k$-truncated SVD process by
resetting those elements back to the known values at each step.
Referring to Eq. (3), the Spectrum-based SCR treats the known
ratings as the boundary conditions, and use $k$-truncated SVD as the
matrix operator $\mathfrak{D}$. The iteration will converge to a
stable matrix $\tilde{R}$, namely the predicted matrix.

\begin{figure}
\scalebox{0.6}[0.6]{\includegraphics{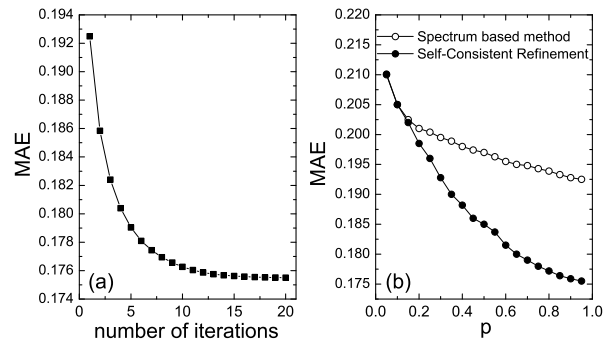}} \caption{(a)
Prediction error vs. iteration step, with $p=0.9$ fixed. (b) The
comparison of algorithmic accuracy between the standard
spectrum-based method and the spectrum-based SCR for different $p$.}
\end{figure}

\emph{Numerical results}.---To test the algorithmic accuracy, we use
a benchmark data set, namely \emph{MovieLens}\cite{data}. The data
consists of $N=3020$ users, $M=1809$ movies, and $2.24\times 10^5$
discrete ratings 1-5. All the ratings are sorted according to their
time stamps. We set a fraction $p$ of earlier ratings as the
training set, and the remain ratings (with later time stamps) as the
testing set.

As shown in Figs. 1 and 2, both the similarity-based and
spectrum-based SCRs converge very fast, and sharply improve the
algorithmic accuracy of the standard methods. In spectrum-based
methods, the parameter $k$ is not observable in the real system,
thus we treat it as a tunable parameter. The results displayed in
Fig. 2 correspond to the optimal $k$ that minimizes the prediction
error. For different $p$, the optimal $k$ is different. Denoting the
data density as $\rho=E/NM$, where $E$ is the number of ratings in
the training set. The spectrum-based SCR will converge only if $k$
is smaller than a threshold \small
\begin{equation}
k_c=\frac{N+M-2}{2}-\sqrt{\left(\frac{N+M-2}{2}\right)^2-NM\rho}\approx
\frac{NM\rho}{N+M-2}.
\end{equation}
\normalsize So that the searching horizon of optimal $k$ can be
reduced to the natural numbers not larger than $k_c$. The
mathematical derivation and numerical results about this threshold
behavior, as well as the sensitivity of algorithmic performance to
$k$ will be discussed elsewhere.

\emph{Conclusions}.---In this Letter, we proposed a algorithmic
framework for recommender systems, namely self-consistent
refinement. This general framework is implemented by embedding two
representative recommendation algorithms: similarity-based and
spectrum-based methods. Numerical simulations on a benchmark data
set demonstrate the significant improvement of algorithmic accuracy
compared with the standard algorithms. Actually, the spectrum-based
SCR has higher accuracy than the similarity-based one, but it
requires an optimizing process on the selection of the parameter
$k$, thus takes longer computational time.

Besides the similarity-based and spectrum-based methods, very
recently, some new kinds of recommendation algorithms that mimic
certain physics dynamics, such as heat conduction \cite{Zhang2007a}
and mass diffusion \cite{Zhang2007b}, are suggested to be the
promising candidates in the next generation of recommender systems
for they provide better algorithmic accuracy while have lower
computational complexity. It is worthwhile to emphasize that those
two algorithms \cite{Zhang2007a,Zhang2007b} also belong to the
framework of SCR - they are just two specific realizations of SCR if
considering the matrix operator $\mathfrak{D}$ as the conduction of
heat or the exchange of mass during one step. In fact, the SCR
framework is of great generality, and any algorithm that can be
expressed in the form of Eq. (2) has the opportunity being improved
via iterative SCR. Furthermore, the present method can be applied in
not only the recommender systems, but also many other subjects, such
as data clustering, miss data mining, detection of community
structure, pattern recognition, predicting of protein structure, and
so on.

This work is partially supported by SBF (Switzerland) for financial
support through project C05.0148 (Physics of Risk), and the Swiss
National Science Foundation (205120-113842). T.Z. acknowledges NNSFC
under Grant No. 10635040 and 60744003, as well as the 973 Project
2006CB705500.


\begin{thebibliography}{1}
\bibitem{norm} The Fribenius norm (also called Euclidean norm, Schui norm or Hilbert-Schmidt norm) of a matrix $\{a_{ij}\}$, is defined as $\|A\|=\left( \sum_i\sum_j a_{ij}^2\right)^{1/2}$.
\end{thebibliography}
\end{document}